\documentclass[12pt]{article}
\usepackage{latexsym,graphicx,amssymb,amsmath}

\def\bseq{\begin{subequation}}  
\def\eseq{\end{subequation}}
\def\bsea{\begin{subeqnarray}}  
\def\esea{\end{subeqnarray}}

\newcommand{\bbox}{\lower.2ex\hbox{$\Box$}}

\newcommand{\beq}{\begin{equation}}
\newcommand{\eeq}{\end{equation}}
\newcommand{\bea}{\begin{eqnarray}}
\newcommand{\eea}{\end{eqnarray}}
\newcommand{\ena}{\end{eqnarray}}
\newcommand{\ba}{\begin{array}}
\newcommand{\ea}{\end{array}}
\newcommand{\ben}{\begin{enumerate}}
\newcommand{\een}{\end{enumerate}}
\newcommand{\bde}{\begin{description}}
\newcommand{\ede}{\end{description}}

\newcommand{\pa}{\partial}

\newcommand{\p}{\pi}

\newcommand{\q}{\quad}

\begin{document}
\begin{titlepage}
\begin{flushright}
IFUM-834-FT\\
Bicocca-FT-05-10\\
June 2005
\end{flushright}
\vspace{2cm}

\noindent{\Large \bf On ${\cal N}=1$ exact superpotentials from
$U(N)$ matrix models}
\vspace{.5cm}
{\bf \hrule width 16.cm}
\vspace {1cm}

\noindent{\large \bf Federico Elmetti$^{a}$, Alberto Santambrogio$^{b}$ and
Daniela Zanon$^{a}$}

\vskip 2mm

{\small \noindent $^a$ Dipartimento di Fisica, Universit\`a di Milano and
 INFN, Sezione di Milano, Via Celoria 16, I-20133 Milano, Italy}

{\small \noindent $^b$ Dipartimento di Fisica, Universit\`a di
Milano-Bicocca and
 INFN, Sezione di Milano, Piazza della Scienza 3, I-20126 Milano, Italy}

\vfill
\begin{center}
{\bf Abstract}
\end{center}
{\small In this letter we compute the exact effective superpotential of ${\cal N}=1$ $U(N)$
supersymmetric gauge theories with $N_{f}$ fundamental flavors and an \emph{arbitrary} tree-level
polynomial superpotential for the adjoint Higgs field. We use the matrix model approach in the
maximally confining phase. When restricted to the case of a tree-level even  polynomial
superpotential, our computation reproduces the known result of the $SU(N)$ theory.} \vspace{2mm}
\vfill \hrule width 6.cm
\begin{flushleft}
e-mail: federico.elmetti@mi.infn.it\\
e-mail: alberto.santambrogio@mi.infn.it\\
e-mail: daniela.zanon@mi.infn.it
\end{flushleft}
\end{titlepage}


In a series of papers \cite{DV1,DV2} Dijkgraaf and Vafa  argued
that for a wide class of ${\cal N}=1$ $U(N)$ supersymmetric gauge
theories the effective superpotential, thought as a function of
the chiral glueball superfield $S$, could be computed by means of
a simple matrix model whose action is the tree-level
superpotential of the gauge theory. Their proposal was the result
of a detailed study of the various dualities between $U(N)$
supersymmetric gauge theories, B-model topological strings and
matrix models in the large $N$ limit \cite{V,HV,CIV,DV3,GV}. This
kind of limit was actually an old idea due to 't Hooft \cite{H}
who had shown that  perturbation theory in the large $N$ limit
singles out only the planar Feynman diagrams. What Diikgraaf and
Vafa discovered was that a leading-order perturbative calculation
via a matrix model could capture completely an exact quantity  in
the corresponding gauge theory, namely the effective
superpotential  which describes the effects of gaugino
condensation. In particular they showed that for a theory with a
matter field in the adjoint representation of the gauge group the
effective superpotential is always of the form: 
\beq\label{chi=2}
W_{eff}(S)=N\frac{\partial \mathcal{F}_{\chi=2}(S)}{\partial S}
\eeq 
where $\mathcal{F}_{\chi=2}$ represents the planar free
energy  from diagrams with the topology of the sphere. Then this
result was obtained directly using a perturbative field theory
approach \cite{DGLVZ}, and through the analysis of the generalized
Konishi anomaly \cite{Wittenetal}. Many extensions of this idea
have been studied, with the aim  to include also quark fields in
the (anti)fundamental representation
\cite{NSW1,NSW2,ACFH,Mc,BR,Oo,F,J}. In particular in \cite{ACFH}
it was realized that such a generalization could be achieved by
taking into account the contributions from planar diagrams with
the topology of the disk ($\mathcal{F}_{\chi=1}$), i.e. the
diagrams with one quark-loop boundary \cite{ACFH}. In this case
the relation (\ref{chi=2}) was modified as follows:
\beq\label{chi=1} W_{eff}(S)=N\frac{\partial
\mathcal{F}_{\chi=2}(S)}{\partial S}+\mathcal{F}_{\chi=1}(S) \eeq

The validity of this approach was tested considering a theory
whose lagrangian contains a mass term for the quark fields $q$ $
(\widetilde{q})$, a Yukawa interaction with the field in the
adjoint $\phi$ and a quadratic tree-level superpotential for
$\phi$. This model has been studied also for different gauge
groups \cite{JO,AN,OW} and a generalized Yukawa coupling
\cite{BHR}.

In this letter we focus on a generalization of this kind of models, with a $\mathcal{N}=1$ $U(N)$
gauge theory with $N_{f}$ flavors and an \emph{arbitrary} tree-level polynomial superpotential for
$\phi$, i.e.
\beq\label{superpotgen}
W(\phi)=\sum_{k=1}^{n+1}\frac{1}{k}g_k \phi^k
\eeq
\noindent We want to obtain the matter
contribution $\mathcal{F}_{\chi=1}(S)$ for this general case. Following \cite{KW} we study the
maximally-confining phase of the theory and compute $\mathcal{F}_{\chi=1}$ explicitly  as a power
series in $S$, i.e. \beq \mathcal{F}_{\chi=1}(S)=\sum_{j}P_{j}S^{j}\eeq \noindent We find that the
coefficients $P_{j}$ for given $j$ depend only on the $g_k$ couplings in the potential with $k<2j$.

Our general derivation naturally includes the results obtained by Gomez-Reino \cite{GR}, who solves
a $SU(N)$ matrix model using the properties of the factorization of the Seiberg-Witten curve. This
is a direct consequence of the fact that for a maximally confining $SU(N)$ theory only the moduli
carrying even indices contribute to the factorization of the Seiberg-Witten curve \cite{Fe}.
Therefore restricting our result to the case of an even polynomial tree-level superpotential we
recover the results in \cite{GR}.

\vspace{0.8cm}

We consider a $\mathcal{N}=1$ $U(N)$ supersymmetric gauge theory obtained by softly breaking
$\mathcal{N}=2$ supersymmetry via the introduction of a  tree-level superpotential. The action is
given by 
\beq\label{theory} 
S=S_{matter}+S_{gauge}+S_{break} 
\eeq 
with \bea S_{matter} &=& \int d^{4}x \int
d^{2}\theta ~d^{2}\overline{\theta}
~\Big(e^{-V}\overline{\phi}e^{V}\phi+\overline{q}e^{V}q+\widetilde{q}
e^{V}\overline{\widetilde{q}}\Big)\nonumber\\
S_{gauge} &=& 2\pi i \tau \int d^{4}x
\int d^{2}\theta~ Tr(W^{\alpha}W_{\alpha})\nonumber\\
S_{break} &=& \int d^{4}x \int d^{2}\theta ~Tr\Big(W_{tree}(\phi,q,\widetilde{q})\Big) \eea and
\beq\label{superpot} W_{tree}(\phi,q,\widetilde{q})=W(\phi)+m\widetilde{q}q-\widetilde{q}\phi q
\eeq where $\phi$ is the matter field in the adjoint representation of $U(N)$, $q$ and
$\widetilde{q}$ $N_{f}$ pairs of quark fields (with mass $m$) in the fundamental and
anti-fundamental and $W^{\alpha}$ the field-strength of the theory. The
superpotential $W(\phi)$ has the general form given in (\ref{superpotgen}).

At the quantum level the vacua of the theory are determined by the appearance
of a gaugino condensate described by a chiral superfield
\beq
S=\frac{1}{32\pi^{2}}Tr(W^{\alpha}W_{\alpha})
\label{gaugino}
\eeq
It is the effective superpotential $W_{eff}(S)$ which encodes the vacuum structure of the theory.

We follow the matrix model approach of \cite{DV1,DV2,DV3} with the generalization to include fundamental
matter fields \cite{ACFH}, so that we
replace $\phi$ with a $N\times N$ hermitian
matrix $\Phi^{a}_{~b}$ ($a,b=1,...,N$),$ $ $q$ with  a $N\times
N_{f}$ matrix $Q^{a}_{\alpha}$ and $\widetilde{q}$ with a
$N_{f}\times N$ matrix $\widetilde{Q}_a^{\alpha}$
($\alpha=1,...,N_{f}$) and write the matrix integral
\beq\label{partition}
Z=\frac{1}{Vol(U(N))}\int d\Phi ~dQ ~d\widetilde{Q}
~e^{-\frac{1}{g_{s}}W_{tree}(\Phi,Q,\widetilde{Q})}
\eeq
where $Vol(U(N))$ is the volume of the gauge group  and $g_{s}$ is
the string coupling.

It is well known \cite{ACFH} that in the 't Hooft limit, where we let
$N\rightarrow\infty,$  $g_{s}\rightarrow 0$ while keeping
$Ng_{s}$ fixed, the matrix model partition function $Z$ receives
contributions from planar diagrams both with the topology of the sphere
($\chi=2$) and the topology of the disk ($\chi=1$) corresponding
respectively to 0 and 1 quark boundary. Thus, if we call
$\mathcal{F}$ the free energy of the model, we have
\beq\label{etof}
Z=e^{\mathcal{F}}\approx
e^{\frac{\mathcal{F}_{\chi=2}}{g_{s}^{2}}+\frac{\mathcal{F}_{\chi=1}}{g_{s}}+...}
\eeq

Now, if we interpret $Ng_s=S$ as the glueball chiral superfield of the
gauge theory (\ref{theory}), we are led to the expression in (\ref{chi=1}.)
The first term in (\ref{chi=1}) is exactly the glueball superpotential
conjectured by Dijkgraaf-Vafa for theories with fields only in the
adjoint representation while the second one comes from the extension
 \cite{ACFH} to include  quark fields in the
(anti)fundamental.

We want to evaluate this matter
contribution $\mathcal{F}_{\chi=1}$ for an arbitrary polynomial superpotential for $\Phi$ of the
form (\ref{superpotgen}). As shown in \cite{NSW1} $\mathcal{F}_{\chi=1}$ can
be written in terms of an hyperelliptic curve $y(x)$ as follows:
\beq\label{effe2}
\mathcal{F}_{\chi=1}=-\frac{1}{2}N_{f}\int_{m}^{\Lambda_{0}}(W^{\prime}(x)-y(x))~
dx
\eeq
where $m$ is the mass of the quarks and $\Lambda_{0}$ a regularization cut-off.
 In order to
integrate the hyperelliptic curve $y(x)$ in (\ref{effe2})
here we will follow the approach of \cite{KW}.  We
study the so-called maximally confining
phase of the theory so that the hyperelliptic curve degenerates as
\beq
y(x)=\sqrt{W^{\prime}(x)^{2}-f_{n-1}(x)}=P_{n-1}(x)
\sqrt{(x-\sigma)^{2}-\mu^{2}}
\eeq
where $P_{n-1}(x)$ is a polynomial of degree $n-1$ and $\sigma$ is
a parameter that can be shifted to 0 in a $U(N)$ theory.
The crucial point is to perform the
change of variable
\beq\label{change}
x\rightarrow \frac{\mu}{2}~(\xi+\xi^{-1})
\eeq
in order to expand $W(x)$ and $W^{\prime}(x)$ in series of $\xi$
and $\xi^{-1}$:
\bea
W(x) &=& W\Big(\frac{\mu}{2}~
(\xi+\xi^{-1})\Big)=b_{0}+\sum_{k=1}^{n+1}b_{k}(\xi^{k}+\xi^{-k})
\nonumber\\
W^{\prime}(x) &=& W^{\prime}\Big(\frac{\mu}{2}~
(\xi+\xi^{-1})\Big)=
c_{0}+\sum_{k=1}^{n}c_{k}(\xi^{k}+\xi^{-k})
\label{series}
\ena
In this way one obtains (see \cite{KW} for
details)
\beq
y(x)=P_{n-1}(x)
\sqrt{x^{2}-\mu^{2}}=\sum_{k=1}^{n}c_{k}(\xi^{k}-\xi^{-k})
\eeq
which allows to solve for $S$
\beq\label{esse}
S \equiv \int_{-\mu}^{+\mu}y(x)~dx = \mu \frac{c_{1}}{2}
\eeq
and gives also
\beq
\int y(x)~ dx=-\mu c_{1}log(\xi) + b_{0}+
\sum_{k=1}^{n+1}b_{k}\Big(\xi^{k}-\xi^{-k}\Big)
\eeq

Now we need to compute the coefficients
$b_{k}$ and $c_{k}$ in (\ref{series}). Performing the change of
variable (\ref{change}) we obtain
\beq\label{W}
W(x) = \sum_{j=1}^{n+1}\frac{g_{j}}{j}
\Big(\frac{\mu}{2}\Big)^{j}\Big(\begin{array}{c}
  j \\
  \frac{j}{2} \\
\end{array}\Big) +
\sum_{k=1}^{n+1}\Big(\xi^{k}+\xi^{-k}\Big)\sum_{j=k}^{n+1}
\frac{g_{j}}{j}\Big(\frac{\mu}{2}\Big)^{j}\Big(\begin{array}{c}
  j \\
  \frac{j-k}{2} \\
\end{array}\Big)
\eeq
with the following convention:
\beq
\Big(\begin{array}{c}
  j \\
  \frac{2m+1}{2} \\
\end{array}\Big)\equiv 0
\qquad\qquad
\Big(\begin{array}{c}
  j \\
  2m \\
\end{array}\Big)\equiv \frac{j!}{(2m)!(j-2m)!}
\eeq
From (\ref{series}) and (\ref{W}) we learn that:
\bea\label{coeff}
b_{0} &=& \sum_{j=1}^{n+1}\frac{g_{j}}{j}
\Big(\frac{\mu}{2}\Big)^{j}\Big(\begin{array}{c}
  j \\
  \frac{j}{2} \\
\end{array}\Big)
\nonumber\\
b_{k} &=& \sum_{j=k}^{n+1}
\frac{g_{j}}{j}\Big(\frac{\mu}{2}\Big)^{j}\Big(\begin{array}{c}
  j \\
  \frac{j-k}{2} \\
\end{array}\Big)
\ena 
Following an analogous procedure we obtain the value of
$c_{1}$ which is needed in (\ref{esse}): \beq \label{c1}
c_{1}=\sum_{j=1}^{n}g_{j+1}\Big(\frac{\mu}{2}\Big)^{j}
\Big(\begin{array}{c}
  j \\
  \frac{j-1}{2} \\
\end{array}\Big)
\eeq
As shown in \cite{KW} one finds
\bea\label{effe3}
\mathcal{F}_{\chi=1} &=& 
 -\frac{1}{2}N_{f}\Bigg\{-W(m)+ \mu
c_{1}log\Big(\frac{\xi(\Lambda_{0})}{\xi(m)}\Big)+b_{0}
\nonumber\\
&& +\sum_{k=1}^{n+1}b_{k}\Big(\frac{m}{\mu}\Big)^{k}\Bigg[
\Bigg(1+
\sqrt{1-\Big(\displaystyle\frac{\mu}{m}\Big)^{2}}~\Bigg)^{k}
-\Bigg(1- \sqrt{1-\Big(\frac{\mu}{m}\Big)^{2}}~\Bigg)^{k}
\Bigg]\Bigg\} 
\ena 
where from (\ref{change}) one can see that:
\bea 
\xi(\Lambda_{0}) &=& \frac{\Lambda_{0}}{\mu}~
\Big(1+\sqrt{1-\Big(\frac{\mu}{\Lambda_{0}}\Big)^{2}}~
\Big)\approx \frac{2\Lambda_{0}}{\mu}\qquad \Lambda_0\rightarrow\infty
\nonumber\\
\xi(m) &=& \frac{m}{\mu}~
\Big(1+\sqrt{1-\Big(\frac{\mu}{m}\Big)^{2}}~\Big)
\ena
Using the following identity
\beq
(1+a)^{k}-(1-a)^{k}=2\sum_{m=0}^{[\frac{k-1}{2}]}\Big(\begin{array}{c}
  k \\
  2m+1 \\
\end{array}\Big)~a^{2m+1}
\eeq
we can rewrite (\ref{effe3}) as
\bea\label{effe4}
\mathcal{F}_{\chi=1} &=& -\frac{1}{2}N_{f}\Bigg[-W(m)+
S~ log \Big(\frac{2\Lambda_{0}}{m}\Big)+b_{0}-2S~
log \Big(\frac{1+\sqrt{1-(\mu/m)^{2}}}{2}\Big)
\nonumber\\
&&~~~~~~+2\sum_{k=1}^{n+1}b_{k}\Big(\frac{m}{\mu}\Big)^{k}
\sum_{l=0}^{[\frac{k-1}{2}]}\Big(\begin{array}{c}
  k \\
  2l+1 \\
\end{array}\Big)~\Big(1-\Big(\frac{\mu}{m}\Big)^{2}\Big)
^{\frac{2l+1}{2}}\Bigg]
\ena
This equation can be easily expanded in
series of $\mu^{2}$.
Using
\bea\label{expans}
log~\Big(\frac{1+\sqrt{1-(\mu/m)^{2}}}{2}\Big) &=& -\sum_{j\geq
1}\frac{1}{2j} \frac{1}{2^{2j}}
\Big(\begin{array}{c}
  2j \\
  j\\
\end{array}\Big)\Big(\frac{\mu}{m}\Big)^{2j}
\nonumber\\
\Big(1-\Big(\frac{\mu}{m}\Big)^{2}\Big)
^{\frac{2l+1}{2}} &=& \sum_{j\geq
0}(-)^{j}\Big(\begin{array}{c}
  \frac{2l+1}{2} \\
  j\\
\end{array}\Big)\Big(\frac{\mu}{m}\Big)^{2j}
\ena
and the expansion in
(\ref{coeff}), finally we obtain
\bea\label{effe5}
\mathcal{F}_{\chi=1} &=& -\frac{1}{2}N_{f}\Bigg[-\sum_{j=1}^{n+1}
\frac{g_{j}}{j}m^{j}+ S~ log~
\Big(\frac{2\Lambda_{0}}{m}\Big)+
\sum_{j=1}^{[\frac{n+1}{2}]}\frac{g_{2j}}{2j}
\Big(\frac{\mu}{2}\Big)^{2j}\Big(\begin{array}{c}
  2j \\
  j \\
\end{array}\Big)
\nonumber\\
&& +\Bigg(\sum_{j=1}^{[\frac{n+1}{2}]}g_{2j}
\Big(\frac{\mu}{2}\Big)^{2j}\Big(\begin{array}{c}
  2j \\
  j \\
\end{array}\Big)\Bigg)~ \Bigg(\sum_{j\geq
1}\frac{1}{2j} \frac{1}{2^{2j}}
\Big(\begin{array}{c}
  2j \\
  j\\
\end{array}\Big)\Big(\frac{\mu}{m}\Big)^{2j}\Bigg)
\nonumber\\
&& +2\sum_{k=1}^{n+1}\Bigg( \sum_{i=k}^{n+1}
\frac{g_{i}}{i}\Big(\frac{\mu}{2}\Big)^{i}\Big(\begin{array}{c}
  i \\
  \frac{i-k}{2} \\
\end{array}\Big) \Bigg)
\sum_{l=0}^{[\frac{k-1}{2}]}\Big(\begin{array}{c}
  k \\
  2l+1 \\
\end{array}\Big)\Bigg(\sum_{j\geq
0}(-)^{j}\Big(\begin{array}{c}
  \frac{2l+1}{2} \\
  j\\
\end{array}\Big)\Big(\frac{\mu}{m}\Big)^{2j-k}\Bigg)\Bigg]
\nonumber\\
\eea
Our final aim is to write $\mathcal{F}_{\chi=1}$ as a power series in $S$. In order to
do so we have to work on (\ref{effe5}) showing that it can be drastically simplified.

\vspace{0.8cm}

Let us consider a matrix model with $N_{f}$ flavors and a general superpotential
for the chiral superfield in the adjoint representation of $U(N)$, given by the
sum of an even polynomial $W_{2n}$ and an odd one $W_{2n+1}$ :
\beq\label{evenpot}
W(x)=W_{2n}(x)+W_{2n+1}(x)=\sum_{j=1}^{n}\frac{g_{2j}}{2j}x^{2j}
+\sum_{j=1}^{n}\frac{g_{2j+1}}{2j+1}x^{2j+1} \eeq
\noindent
We notice that,
as explained in \cite{Fe}, the case $W(x)=W_{2n}(x)$ corresponds
to a $SU(N)$ theory (for which only the even terms contribute to
the glueball superpotential). The $SU(N)$ theory has been considered in
\cite{GR}, where the effective superpotential was computed order by
order in $S$ using factorization properties of the
Seiberg-Witten curve. Solving the general case in (\ref{evenpot}) we will be able to compare
our results with those in \cite{GR}.

First we rewrite (\ref{effe5}) separating the even part from
the odd one:
\bea\label{effe6} &&
\mathcal{F}_{\chi=1}=-\frac{1}{2}N_{f}\Bigg[-\sum_{j=1}^{n}
\frac{g_{2j}}{2j}m^{2j}-\sum_{j=1}^{n}\frac{g_{2j+1}}{2j+1}m^{2j+1}+
S~log~\Big(\frac{2\Lambda_{0}}{m}\Big)+
\nonumber\\
&& 
+ \sum_{j=1}^{n}\frac{g_{2j}}{2j}
\Big(\frac{\mu}{2}\Big)^{2j}\Big(\begin{array}{c}
  2j \\
  j \\
\end{array}\Big)\,+
\Bigg(\sum_{i=1}^{n}g_{2i}
\Big(\frac{\mu}{2}\Big)^{2i}\Big(\begin{array}{c}
  2i \\
  i \\
\end{array}\Big)\Bigg)~\Bigg(\sum_{j\geq
1}\frac{1}{2j} \frac{1}{2^{2j}} \Big(\begin{array}{c}
  2j \\
  j\\
\end{array}\Big)\Big(\frac{\mu}{m}\Big)^{2j}\Bigg)\,+
\nonumber\\
&& +2\sum_{k=1}^{n}\Bigg(
\sum_{i=k}^{n}\frac{g_{2i}}{2i}\Big(\frac{\mu}{2}\Big)^{2i}\Big(\begin{array}{c}
  2i \\
  i-k \\
\end{array}\Big) \Bigg)
\sum_{l=0}^{k-1}\Big(\begin{array}{c}
  2k \\
  2l+1 \\
\end{array}\Big)\Bigg(\sum_{j\geq
0}(-)^{j}\Big(\begin{array}{c}
  \frac{2l+1}{2} \\
  j\\
\end{array}\Big)\Big(\frac{\mu}{m}\Big)^{2(j-k)}\Bigg)\,+
\nonumber\\
&& +\sum_{k=0}^{n}\Bigg(
\sum_{i=k}^{n}\frac{g_{2i+1}}{2i+1}\Big(\frac{1}{2}\Big)^{2i}
\Big(\begin{array}{c}
  2i+1 \\
  i-k \\
\end{array}\Big) \Bigg)
\sum_{l=0}^{k}\Big(\begin{array}{c}
  2k+1 \\
  2l+1 \\
\end{array}\Big)
\nonumber\\
&&~~~~~~~~~~~~~~~~~~~~~~~~~~~~~~~~~~~~~~~~~~~~~
\Bigg(\sum_{j\geq
0}(-)^{j}\Big(\begin{array}{c}
  \frac{2l+1}{2} \\
  j\\
\end{array}\Big)m^{2k-2j+1}\mu^{2(j-k+i)}\Bigg)
\Bigg]
\eea
\noindent
In order to obtain  $\mathcal{F}_{\chi=1}$ as a power series of $S$ we proceed 
in two steps. First we look for an expansion in $\mu^2$. Then we 
will express $\mu^2$ as a power series in  $S$ by means of the formulas 
(\ref{esse}) and (\ref{c1}).
\noindent
To this end we organize eq. (\ref{effe6}) in the following way:
\bea\label{effeABC} 
&&\mathcal{F}_{\chi=1}=-\frac{1}{2}N_{f}\Bigg[-\sum_{j=1}^{n}
\frac{g_{2j}}{2j}m^{2j}-\sum_{j=1}^{n}\frac{g_{2j+1}}{2j+1}m^{2j+1}+
S~log~\Big(\frac{2\Lambda_{0}}{m}\Big)+
\nonumber\\
&&\qquad\qquad\qquad
+\sum_{j=1}^{n}A_{j}(\mu^{2})^{j}+\sum_{j\geq
2}B_{j}(\mu^{2})^{j}+\sum_{j\geq 0}C_{j}(\mu^{2})^{j}+
\sum_{j\geq 0}\widetilde{C}_{j}(\mu^{2})^{j}\Bigg] 
\ena 
where the seven terms correspond to the seven terms in (\ref{effe6}).
\\
The $A_j$ coefficients are immediately identified
\beq
A_{j}\equiv\frac{g_{2j}}{2j} \frac{1}{2^{2j}}\Big(\begin{array}{c}
  2j \\
  j \\
\end{array}\Big).
\eeq
The $B_j$ coefficients are not too difficult to compute and one finds
\beq\label{BB}
B_{j}\equiv
\sum_{\substack{i=1\\i\leq n}}^{j-1}
\frac{g_{2i}}{2^{2j}}\Big(\begin{array}{c}
  2i \\
  i \\
\end{array}\Big)\Big(\begin{array}{c}
  2(j-i) \\
  j-i\\
\end{array}\Big)\frac{1}{2(j-i)m^{2(j-i)}}
\eeq 
On the other hand the computation of $C_j$ and 
$\widetilde{C}_j$ is a more 
difficult task. Let us concentrate on them.
In the last two terms in (\ref{effe6}) we
perform the change of index $i\rightarrow i+k$ 
and a shift
%
$j\rightarrow j-i$. In this way we obtain
\bea\label{coeffC4} 
C_{j}&\equiv&\sum_{k=1}^{n}
\sum_{\substack{i=k\\i\leq
n}}^{j+k}\frac{g_{2i}}{2i}\Big(\begin{array}{c}
  2i \\
  i-k \\
\end{array}\Big) \frac{1}{2^{2i-1}}\frac{(-)^{j-i+k}}{m^{2(j-i)}}
\Bigg[\sum_{l=0}^{k-1}\Big(\begin{array}{c}
  2k \\
  2l+1 \\
\end{array}\Big)\Big(\begin{array}{c}
  \frac{2l+1}{2} \\
  j-i+k\\
\end{array}\Big)\Bigg]
\\\label{Cterm4}
\widetilde{C}_{j}&\equiv&\sum_{k=0}^{n}
\sum_{\substack{i=k\\i\leq
n}}^{j+k}\frac{g_{2i+1}}{2i+1}\Big(\begin{array}{c}
  2i+1 \\
  i-k \\
\end{array}\Big) \frac{(-)^{j-i+k}}{2^{2i}m^{2(j-i)-1}}
\Bigg[\sum_{l=0}^{k}\Big(\begin{array}{c}
  2k+1 \\
  2l+1 \\
\end{array}\Big)\Big(\begin{array}{c}
  \frac{2l+1}{2} \\
  j-i+k\\
\end{array}\Big)\Bigg]
\ena
The sum over $l$ in (\ref{coeffC4}) and
(\ref{Cterm4}) can be rewritten in a nice form: \beq\label{somma1}
\sum_{l=0}^{k-1}\Big(\begin{array}{c}
  2k \\
  2l+1 \\
\end{array}\Big)\Big(\begin{array}{c}
  \frac{2l+1}{2} \\
  j-s+k\\
\end{array}\Big)\equiv 2^{2(s-j)}\frac{k}{(j-s+k)!}
~\frac{\Gamma(k+s-j)}{\Gamma(2s-2j+1)} \eeq

\beq \label{somma2}\sum_{l=0}^{k-1}\Big(\begin{array}{c}
  2k+1 \\
  2l+1 \\
\end{array}\Big)\Big(\begin{array}{c}
  \frac{2l+1}{2} \\
  j-s+k\\
\end{array}\Big)\equiv \frac{2^{-2k-1}\sqrt{\pi}}
{(j-s+k)!(2k)!}
~\frac{\Gamma(2k+2)\,\Gamma(2j-2s-1)}{\Gamma(k-j+s+\frac{3}{2})
\,\Gamma(2j-2s-2k-1)}
\eeq

We start computing the lowest terms: \bea\label{CC1} C_{0} &=&
\sum_{k=1}^{n}\frac{m^{2k}}{2^{2k-1}} \frac{g_{2k}}{2k}
\sum_{l=0}^{k-1}\Big(\begin{array}{c}
  2k \\
  2l+1 \\
\end{array}\Big)= \sum_{k=1}^{n}m^{2k}
\frac{g_{2k}}{2k}\nonumber\\
\label{Clowest1} \widetilde{C}_{0} &=&
\sum_{k=0}^{n}\frac{m^{2k+1}}{2^{2k}} \frac{g_{2k+1}}{2k+1}
\sum_{l=0}^{k}\Big(\begin{array}{c}
  2k+1 \\
  2l+1 \\
\end{array}\Big)= \sum_{k=0}^{n}m^{2k+1}
\frac{g_{2k+1}}{2k+1}\eea
In fact these coefficients are exactly
cancelled by the first two terms in
(\ref{effeABC}). Moreover we want to show that all the other
coefficients $C_{j}$ and $\widetilde{C}_{j}$, $j\neq 0$ receive
contributions only from the first $2j$ and $2j-1$ terms
of the polynomial superpotential respectively. 

Therefore we need to show that every term in
(\ref{coeffC4}) and (\ref{Cterm4}) with an index $i>j$ does not contribute.
Let us consider $i=s$ for some
$s>j$. It is not hard to see that 
the contribution in (\ref{coeffC4}) proportional to $g_{2s}$ can be written as
a sum over
$k=s-j,...,s$ 
\beq\label{full}
g_{2s}\Bigg[\sum_{k=s-j}^{s}\frac{1}{s} \Big(\begin{array}{c}
  2s \\
  s-k \\
\end{array}\Big) \frac{1}{2^{2s}}\frac{(-)^{j-s+k}}{m^{2(j-s)}}
\Bigg(\sum_{l=0}^{k-1}\Big(\begin{array}{c}
  2k \\
  2l+1 \\
\end{array}\Big)\Big(\begin{array}{c}
  \frac{2l+1}{2} \\
  j-s+k\\
\end{array}\Big)\Bigg)\Bigg]
\eeq
In the same way the contribution in (\ref{Cterm4}) proportional to $g_{2s+1}$ 
becomes
\beq\label{g2s+1} 
g_{2s+1}\Bigg[\sum_{k=s-j}^{s}\frac{1}{2s+1}
\Big(\begin{array}{c}
  2s+1 \\
  s-k \\
\end{array}\Big) \frac{1}{2^{2s}}\frac{(-)^{j-s+k}}{m^{2(j-s)-1}}
\Bigg(\sum_{l=0}^{k-1}\Big(\begin{array}{c}
  2k+1 \\
  2l+1 \\
\end{array}\Big)\Big(\begin{array}{c}
  \frac{2l+1}{2} \\
  j-s+k\\
\end{array}\Big)\Bigg)\Bigg]
\eeq
\noindent
Now, setting $t\equiv s-j$, $t>0$ (\ref{full}) becomes proportional to
\bea
&&\sum_{k=0}^{j}\Big(\begin{array}{c}
  2(j+t) \\
  j-k \\
\end{array}\Big) (-)^{k}(k+t)~
\frac{\Gamma(k+2t)}{\Gamma(k+1)}=
\nonumber\\
&& =\frac{4^{j}\,t \, \Gamma(j+1/2)\,
\Gamma(2t)}{\sqrt{\pi}\, j \, \Gamma(j)}- \frac{\Gamma(2j+1)\,
\Gamma(2t+1)}{2\, j^{2}\, \Gamma^{2}(j)}=
\nonumber\\
&& =\frac{2\,j\,t \, \Gamma(2t)}{\sqrt{\pi}\,j^{2}\,
\Gamma^{2}(j)}\Big(2^{2j-1}\,\Gamma(j)\,\Gamma(j+1/2)-
\sqrt{\pi}\,\Gamma(2j) \Big)\equiv 0 \ena
\noindent
In a similar manner from (\ref{g2s+1}) we obtain
\beq \sum_{k=s-j}^{s}\frac{(-)^{k}}{2^{2k}}
\frac{\Gamma(2k+2)}{(j-s+k)!\,\Gamma(k-j+s+\frac{3}{2})}\,
\frac{\Gamma(2j-2s-1)}{\Gamma(2j-2s-2k-1)(2k)!}\eeq
which is proportional to
\beq
4^{j}\,\Gamma(j+\frac{1}{2})\,\Gamma^{2}(j+1)-2\Gamma(j)\Big[j\,\sqrt{\pi}\,(j-2s-2)\,
\Gamma(2j)+4^{j}(1+s)\,\Gamma(j+\frac{1}{2})\,\Gamma(j+1)\Big]\equiv
0 \eeq
We have used repeatedly the following property of the $\Gamma$ matrices (for
integer $j$): \beq\label{gamma}
\Gamma(j+1/2)\equiv\frac{\sqrt{\pi}}{2^{2j-1}}
\,\frac{\Gamma(2j)}{\Gamma(j)} \eeq 
This completes our proof.

\vspace{0.6cm}

At this point we are left with: 
\beq C_{j}=\sum_{s=1}^{j}c_{js}g_{2s}
\,\,\,\,\,\,\,\,\,\,\,\,\,\,\,\,\,\,\,\,\,\,\,
\widetilde{C}_{j}=\sum_{s=1}^{j-1}\widetilde{c}_{js}g_{2s+1}\nonumber\eeq
where
\beq c_{js}\equiv \sum_{k=1}^{s}\frac{(-)^{j-s+k}}{2^{2s}s\,
m^{2(j-s)}}\Big(\begin{array}{c}
  2s \\
  s-k \\
\end{array}\Big)\sum_{l=0}^{k-1}\Big(\begin{array}{c}
  2k \\
  2l+1 \\
\end{array}\Big)\Big(\begin{array}{c}
  \frac{2l+1}{2} \\
  j-s+k\\
\end{array}\Big)\eeq
and
\beq \widetilde{c}_{js}\equiv
\sum_{k=1}^{s}\frac{(-)^{j-s+k}}{2^{2s}(2s+1)
m^{2(j-s)}}\Big(\begin{array}{c}
  2s+1 \\
  s-k \\
\end{array}\Big)\sum_{l=0}^{k}\Big(\begin{array}{c}
  2k+1 \\
  2l+1 \\
\end{array}\Big)\Big(\begin{array}{c}
  \frac{2l+1}{2} \\
  j-s+k\\
\end{array}\Big)
\eeq

Using (\ref{somma1}) and (\ref{somma2}) we find: 
\bea\label{cjj}
&&c_{jj}=-\frac{1}{2^{2j}2j}\Big(\begin{array}{c}
  2j \\
  j\\
\end{array}\Big)
\nonumber\\
&& c_{js}=-\frac{1}{2^{2j-1}j\,s\,m^{2(j-s)}}\,\frac{\Gamma(2s)}{\Gamma^{2}(s)}
\,\Big(\begin{array}{c}
  2(j-s)-1 \\
  j-s\\
\end{array}\Big)\,\,\,\,\,\,\,\,\,\,\,s<j
\nonumber\\
&&\widetilde{c}_{js}=-\frac{1}{2^{2j-1}j\,s^{2}\,m^{2(j-s)-1}}\,\frac{\Gamma(2s+1)}{\Gamma^{2}(s)}
\,\Big(\begin{array}{c}
  2(j-s)-2 \\
  j-s-1\\
\end{array}\Big)\,\,\,\,\,\,\,\,\,\,\,s<j
\ena
\noindent
Note that the $A_{j}$ terms are exactly cancelled by $c_{jj}$ for
$j=1,...n$ computed in (\ref{cjj}), leaving no term linear in $\mu^{2}$ (then
no term linear in $S$
except for the standard piece $S~log(2\Lambda_{0}/m)$). 
So we can rewrite (\ref{effeABC}) as follows:
\beq\label{effe9}
\mathcal{F}_{\chi=1}=-\frac{1}{2}N_{f}\Bigg[S~log~
\Big(\frac{2\Lambda_{0}}{m}\Big)+ \sum_{j\geq 2}(B_{j}+C^{n}_{j}+
\widetilde{C}_{j})(\mu^{2})^{j}\Bigg] \eeq 
where we have defined
\beq
C^{n}_{j}\equiv\left\{ \begin{array}{c} 
C_{j}-g_{2j}c_{jj}\qquad j\leq n \\
C_{j}\qquad\qquad~~~~~ j> n
\end{array}\right.
\eeq
\noindent
We give explicitly the form of the lowest terms for the coefficients $B_j$:
\bea B_{2} &=&
+\frac{1}{8}\frac{g_{2}}{m^{2}}
\nonumber\\
B_{3} &=& +\frac{3}{64}\frac{g_{2}}{m^{4}}+
\frac{3}{32}\frac{g_{4}}{m^{2}}
\nonumber\\
B_{4} &=& +\frac{5}{192}\frac{g_{2}}{m^{6}}+
\frac{9}{256}\frac{g_{4}}{m^{4}} +\frac{5}{64}\frac{g_{6}}{m^{2}}
\ena
and for the coefficients $C_j$ and $\widetilde{C}_j$:
\bea\label{CC2}
C_{1} &=& -\frac{g_{2}}{4}
\nonumber\\
C_{2} &=& -\frac{g_{2}}{16m^{2}} -\frac{3}{32}g_{4}
\nonumber\\
C_{3} &=& -\frac{3}{96}\frac{g_{2}}{m^{4}}-
\frac{3}{96}\frac{g_{4}}{m^{2}} -\frac{5}{96}g_{6}
\nonumber\\
C_{4} &=& -\frac{5}{256}\frac{g_{2}}{m^{6}}-
\frac{9}{512}\frac{g_{4}}{m^{4}}
-\frac{5}{256}\frac{g_{6}}{m^{2}}-\frac{35}{1024}g_{8} \ena
and
\bea\label{Clowest2} \widetilde{C}_{1} &=& 0
\nonumber\\
\widetilde{C}_{2} &=& -\frac{g_{3}}{8 m}
\nonumber\\
\widetilde{C}_{3} &=& -\frac{g_{3}}{24 m^{3}}- \frac{g_{5}}{16 m}
\nonumber\\
\widetilde{C}_{4} &=& -\frac{3}{128}\frac{g_{3}}{m^{5}}-
\frac{3}{128}\frac{g_{5}}{m^{3}} -\frac{5}{128}\frac{g_{7}}{m}
\ena

As a final step we want to reexpress $\mathcal{F}_{\chi=1}$ in a power series of $S$.
From (\ref{esse}) and (\ref{c1}) we know that
\beq\label{essemu}
S=\frac{1}{2}\sum_{j=1}^{[\frac{n+1}{2}]}g_{2j}
\Big(\frac{\mu}{2}\Big)^{2j}\Big(\begin{array}{c}
  2j \\
  j \\
\end{array}\Big)
\eeq
We want to invert this relation in order to obtain $\mu^{2}$ in
terms of $S$.
We look for an expression of the form
\beq\label{muesse}
\mu^{2}=\sum_{m>0}a_{m}S^{m}
\eeq
where $a_{m}$ are the coefficients to be determined. Inserting (\ref{muesse}) into (\ref{essemu}) we obtain
\bea
S &=& \frac{1}{2}\sum_{j=1}^
{[\frac{n+1}{2}]}\frac{g_{2j}}{2^{2j}}
\Big(\begin{array}{c}
  2j \\
  j \\
\end{array}\Big)\Big(\sum_{m>0}a_{m}S^{m}\Big)^{j}
\nonumber\\
&=& \frac{1}{2}\sum_{j=1}^{[\frac{n+1}{2}]}\frac{g_{2j}}{2^{2j}}
\Big(\begin{array}{c}
  2j \\
  j \\
\end{array}\Big)\sum_{\substack{p_{1},p_{2},...=0\\p_{1}+p_{2}+...=j}}^{j}
M(j;p_{1},p_{2},...)a_{1}^{p_{1}}a_{2}^{p_{2}}...S^{p_{1}+2p_{2}+3p_{3}+...}
\ena
where we have defined the coefficient:
\beq
M(j;p_{1},p_{2},...,p_{t},...)\equiv
\frac{j!}{p_{1}!p_{2}!...p_{t}!...}
\eeq
With a bit of labor one can argue that the general
coefficient $a_{m} $ can be expressed recursively in terms of
$a_{1},a_{2},...,a_{m-1}$: 
\bea\label{a}
a_1 &=& \frac{4}{g_2}\nonumber\\
a_{m} &=& -\sum_{\substack{p_{1},...,p_{m-1}=0\\
p_{1}+2p_{2}+...+(m-1)p_{m-1}=m\\p_{1}+p_{2}+...+p_{m-1}\leq
[\frac{n+1}{2}]}}^{m} \frac{2}{g_{2}}~
\Bigg(\begin{array}{c}
  2(p_{1}+...+p_{m-1}) \\
  p_{1}+...+p_{m-1} \\
\end{array}\Bigg)~a_{1}^{p_{1}}...\,a_{m-1}^{p_{m-1}}~ \cdot
\nonumber\\
&& \qquad\qquad
\cdot~\frac{(p_{1}+...+p_{m-1})!}{(p_{1})!...(p_{m-1})!}~
\frac{g_{2(p_{1}+...+p_{m-1})}}{2^{2(p_{1}+...+p_{m-1})}}
\eea
Here we give the expressions of the first coefficients:
\bea\label{aa}
a_{2} &=& -12\frac{g_{4}}{g_{2}^{3}}
\nonumber\\
a_{3} &=& +72\frac{g_{4}^{2}}{g_{2}^{5}}
-40\frac{g_{6}}{g_{2}^{4}}
\nonumber\\
a_{4} &=& -540\frac{g_{4}^{3}}{g_{2}^{7}}
+600\frac{g_{4}g_{6}}{g_{2}^{6}}-140\frac{g_{8}}{g_{2}^{5}}
\nonumber\\
a_{5} &=& +4536\frac{g_{4}^{4}}{g_{2}^{9}}
-7560\frac{g_{4}^{2}g_{6}}{g_{2}^{8}}
+2520\frac{g_{4}g_{8}}{g_{2}^{7}}
+1200\frac{g_{6}^{2}}{g_{2}^{7}}
-504\frac{g_{10}}{g_{2}^{6}}
\ena
Now, using (\ref{muesse}) and (\ref{a}) we are able to give the
expansion of $\mathcal{F}_{\chi=1}$ in series of $S$:

\beq\label{effe10}
\mathcal{F}_{\chi=1}=-\frac{1}{2}N_{f}\Bigg[S~log~
\Big(\frac{2\Lambda_{0}}{m}\Big)+ \sum_{k\geq
2}\Bigg(\sum_{j=2}^{k}(B_{j}+C^{n}_{j}+
\widetilde{C}_{j})\sum_{\substack{
p_{1}...p_{k}=0\\p_{1}+...+p_{k}=j\\p_{1}+...+kp_{k}=k}}^{j}M(j;p_{1},...,p_{k})\,
a_{1}^{p_{1}}...\,a_{k}^{p_{k}}\Bigg)S^{k}\Bigg] \nonumber\eeq

\bea \label{effe11}&=& -N_{f}\Bigg[\frac{1}{2}S~ log~
\Big(\frac{2\Lambda_{0}}{m}\Big)+
\Big(\frac{1}{2g_{2}m^{2}}-\frac{g_{3}}{g_{2}^{2}m}\Big)S^{2}
\nonumber\\
&&+
\Big(\frac{1}{2g_{2}^{2}m^{4}}-\frac{4}{3}\frac{g_{3}}{g_{2}^{3}m^{3}}-
\frac{g_{4}}{g_{2}^{3}m^{2}}+6\frac{g_{3}g_{4}}{g_{2}^{4}m}-2\frac{g_{5}}{g_{2}^{3}m}
\Big)S^{3}
\nonumber\\
&&+ \Big(\frac{5}{6}\frac{1}{g_{2}^{3}m^{6}}-
3\frac{g_{3}}{g_{2}^{4}m^{5}}+
\frac{9}{2}\frac{g_{4}^{2}}{g_{2}^{5}m^{2}}-
\frac{9}{4}\frac{g_{4}}{g_{2}^{4}m^{4}}+12\frac{g_{3}g_{4}}{g_{2}^{5}m^{3}}
-45\frac{g_{3}g_{4}^{2}}{g_{2}^{6}m}
\nonumber\\
&&+18\frac{g_{4}g_{5}}{g_{2}^{5}m}
-3\frac{g_{5}}{g_{2}^{4}m^{3}}-
\frac{5}{2}\frac{g_{6}}{g_{2}^{4}m^{2}} + 20\frac{g_{3}g_{6}}{g_{2}^{5}m}
-5\frac{g_{7}}{g_{2}^{4}m}\Big)S^{4}+...\Bigg] \ena
\noindent
If we extract from (\ref{effe11}) the even contributions we
reproduce exactly what has been computed in \cite{GR}.

\vspace{0.8cm}

In order to have the full expression of the glueball
superpotential (\ref{chi=1}) we have to add the $\chi=2$
contribution to (\ref{effe11}). One can find the implicit solution
to this problem in \cite{Fe} where the  $\chi=2$ contribution is
given in terms of some parameters which are non-linear
functions of $S$. After some algebra we found that the explicit
solution can be written as
\\
\beq\label{Wchi=2} 
N\frac{\pa \mathcal{F}_{\chi=2}}{\pa S}=-NS\, log\Big(\frac{S}{g_{2}\Lambda_{0}^{2}}\Big)-N \sum_{j>0} (W_{j}- \widetilde{W}_{j}) S^{j}\eeq\\
where 
\beq 
W_{j}\equiv\sum_{m=1}^{j-1} \frac{(-)^{m+1}}{m
D_{1}^{m}}\sum_{\substack{p_{1},...,p_{j-1}=0\\p_{1}+...+p_{j-1}=m\\p_{1}+...+(j-1)p_{j-1}=j-1
}}^{m} M(m;p_{1},...,p_{j-1}) D_{2}^{p_{1}}...D_{j}^{p_{j-1}}
\eeq
and
\bea
\widetilde{W}_{j}\equiv\sum_{p=2}^{2j}\frac{g_{p}}{p}\sum_{q=0}^{[p/2]}\Big(\begin{array}{c}
  p \\
  2q \\
\end{array}\Big) \Big(\begin{array}{c}
  2q \\
  q \\
\end{array}\Big) \sum_{\substack{p_{1},...,p_{j-1}=0\\p_{1}+...+p_{j-1}=p-2q
}}^{p-2q} M(p-2q;p_{1},...,p_{j-1})
z_{1}^{p_{1}}...z_{j-1}^{p_{j-1}} \nonumber\\
\sum_{\substack{q_{1},...,q_{j}=0\\q_{1}+...+q_{j}=q\\(p_{1}+q_{1})+...
+(j-1)(p_{j-1}+q_{j-1})+jq_{j}=j}}^{q} M(q;q_{1},...,q_{j})
D_{1}^{q_{1}}...D_{j}^{q_{j}}
\eea
are defined in terms of some parameters $D_{j}$ and $z_{j}$ which
can be computed recursively 
\beq 
D_{j}\equiv
\sum_{k>0}d_{k}\sum_{\substack{p_{1},...,p_{j}=0\\p_{1}+...+p_{j}=k\\p_{1}+...+jp_{j}=j
}}^{k} M(k;p_{1},...,p_{j}) z_{1}^{p_{1}}...z_{j}^{p_{j}}
\eeq
and
\bea
&&z_{1}=-2\frac{g_{3}}{g_{2}^{2}}\nonumber\\
&&z_{j}\equiv
-\frac{1}{g_{2}d_{1}}\sum_{p=2}^{2j}\frac{g_{p}}{p}\sum_{q=1}^{[p/2]}q\Big(\begin{array}{c}
  p \\
  2q \\
\end{array}\Big) \Big(\begin{array}{c}
  2q \\
  q \\
\end{array}\Big) \sum_{\substack{p_{1},...,p_{j}=0\\p_{1}+...+p_{j}=q
}}^{q} M(q;p_{1},...,p_{j})
d_{1}^{p_{1}}...d_{j}^{p_{j}}
\nonumber\\
&&\sum_{\substack{q_{1},...,q_{j-1}=0\\q_{1}+...+q_{j-1}=p-2q+p_{1}+...+jp_{j}\\
q_{1}+... +(j-1)q_{j-1}=j}}^{p-2q+p_{1}+...+jp_{j}}
M(p-2q+p_{1}+...+jp_{j};q_{1},...,q_{j-1})
z_{1}^{q_{1}}...z_{j-1}^{q_{j-1}}
\ena
\\
The $d_{j}$ coefficients which appear in (69) and (70)
are also given recursively: 
\bea 
&&
d_{1}=-\frac{g_{2}}{2g_{3}}\nonumber\\
&&d_{j}\equiv
-\frac{1}{2g_{3}}\sum_{p=3}^{2j}g_{p+1}\sum_{q=0}^{[p/2]}\Big(\begin{array}{c}
  p \\
  2q \\
\end{array}\Big) \Big(\begin{array}{c}
  2q \\
  q \\
\end{array}\Big)\sum_{\substack{p_{1},...,p_{j-1}=0\\p_{1}+...+p_{j-1}=q\\
p-2q+p_{1}+... +(j-1)p_{j-1}=j}}^{q}M(q;p_{1},...,p_{j-1})\;
d_{1}^{p_{1}}...d_{j-1}^{p_{j-1}}\nonumber\\
&&
\ena
\\
Explicitly to the fourth order in $S$ we obtain 
\bea \label{Wchi=2exp}
N\frac{\pa \mathcal{F}_{\chi=2}}{\pa S}=-NS\Big(log\Big(\frac{S}{g_{2}\Lambda_{0}^{2}}\Big)-1\Big)-N\Bigg[
\Big(2\frac{g_{3}^{2}}{g_{2}^{3}}-\frac{3}{2}\frac{g_{4}}{g_{2}^{2}}\Big)S^{2}+
\Big(\frac{32}{3}\frac{g_{3}^{4}}{g_{2}^{6}}-24\frac{g_{3}^{2}g_{4}}{g_{2}^{5}}+\nonumber\\
+\frac{9}{2}\frac{g_{4}^{2}}{g_{2}^{4}}+12\frac{g_{3}g_{5}}{g_{2}^{4}}-\frac{10}{3}
\frac{g_{6}}{g_{2}^{3}}\Big)S^{3}+\Big(\frac{280}{3}\frac{g_{3}^{6}}{g_{2}^{9}}-
340\frac{g_{3}^{4}g_{4}}{g_{2}^{8}}+270
\frac{g_{3}^{2}g_{4}^{2}}{g_{2}^{7}}-\frac{45}{2}\frac{g_{4}^{3}}{g_{2}^{6}}+\nonumber\\
+200 \frac{g_{3}^{3}g_{5}}{g_{2}^{7}} -180
\frac{g_{3}g_{4}g_{5}}{g_{2}^{6}}+18
\frac{g_{5}^{2}}{g_{2}^{5}}-100\frac{g_{3}^{2}g_{6}}{g_{2}^{6}}+30\frac{g_{4}g_{6}}{g_{2}^{5}}
+40
\frac{g_{3}g_{7}}{g_{2}^{5}}-\frac{35}{4}\frac{g_{8}}{g_{2}^{4}}\Big)S^{4}+...\Bigg]
\eea

Let us notice that the $S^k$ term in this expression depends only on the 
low order couplings $g_3, g_4, \ldots, g_{2k}$, in a way parallel to what
we have found for the case of fundamental matter (see (\ref{effe11})). 
When only an adjoint matter field is present this 
behavior follows immediately from the perturbative analysis in \cite{DGLVZ}. 
There it was found that contributions to the $S^k$ term come from a 
planar diagram with exactly $k$ loops, which can contain
vertices with at most $2k$ external legs.\\ 
What we found in (\ref{effe11}) is the corresponding situation where now
the graphs have an extra loop of fundamental matter.
It is very easy to write down the graphs corresponding to every single
term in the expressions (\ref{effe11}) and (\ref{Wchi=2exp}) since
the factors of $g_2$ and $m$ in the denominators count
the number of adjoint and fundamental propagators respectively.

\vspace{0.6cm}

To summarize, we have computed the effective glueball superpotential for 
a ${\cal N}=1$
$U(N)$ gauge theory with matter in the fundamental ($N_f$ flavors) and a
field in the adjoint with a general polynomial tree-level superpotential.
This has been done by using the technology developed in \cite{KW} for
computing the contribution to the matrix model partition function coming
from diagrams with the disk topology.
The full contribution to the glueball superpotential is given in 
the expressions (\ref{effe11}) and (\ref{Wchi=2}).
\\
Our results generalize the ones obtained in \cite{GR} since as discussed in
\cite{Fe} 
the $SU(N)$ theory considered there is equivalent to a $U(N)$ model with
an even polynomial superpotential for the adjoint field.
\noindent\\
It would be interesting to extend these results to the case of a 
superpotential admitting several vacua.

\vspace{0.8cm}

\noindent
{\bf Acknowledgements}

\noindent
This work has been partially supported by INFN, MURST and the European
Commission RTN program MRTN-CT-2004-005104 in which the authors are associated
to the University of Milano-Bicocca.

\end{document}